\documentclass[prl,floatfix,superscriptaddress,nofootinbib,showpacs,showkeys,preprintnumbers,twocolumn]{revtex4}

\usepackage[dvips]{epsfig}
\usepackage{graphicx}

\begin{document}

\title{Study of narrow baryon resonance decaying into $K^0_s p$ in
$pA$-interactions at $70\ GeV/c$ with SVD-2.}

\author{A. Kubarovsky\footnote[1]{e-mail:alex\_k@hep.sinp.msu.ru}, V. Popov\footnote[6]{e-mail:popov@hep.sinp.msu.ru} and V. Volkov\footnote[11]{e-mail:volkov@silab.sinp.msu.ru}\\
(for the SVD-2 collaboration)}

\affiliation{D.V. Skobeltsyn Institute of
Nuclear Physics, Lomonosov Moscow State University, \\
1/2 Leninskie gory, Moscow, 119992 Russia\\}

\begin{abstract}{The inclusive reaction $p A \rightarrow
pK^0_s + X$ was studied at IHEP accelerator with $70\ GeV$ proton beam
using SVD-2 detector. Two different samples of $K^0_s$, statistically
independent and belonging to different phase space regions were used
in the analyses and a narrow baryon resonance with the mass $M=1523\pm
2(stat.)\pm 3(syst.)\ MeV/c^2$ was observed in both samples of the
data.} 
\end{abstract}

\maketitle

\section{Introduction}

In last three years the observation of a narrow baryon state named
$\Theta^+$ predicted by Diakonov, Petrov and Polyakov\cite{dpp} has
been reported by a large number of experiments in the $nK^+$ or
$K^0_s p$ decay channels. Several experiments, mostly at high
energies, did not confirm the existence of $\Theta^+$. The present
situation and complete list of references to positive and null
results can be found in the reviews\cite{dzierba,hicks,danilov}. Here
we present a new study of the reaction $pN\rightarrow \Theta^+ + X$,
\ $\Theta^+ \rightarrow pK^0_s$, $K^0_s \rightarrow \pi^+\pi^-$, with
two independent samples of $K^0_s$  used\cite{svd1}.

\section{The SVD-2 apparatus} A detailed description of SVD-2 detector
and its trigger system can be found elsewhere\cite{svd4,svdtrig}. The
components of the detector used in current analyses are: the
high-precision microstrip vertex detector with active(Si) and
passive(C,Pb) nuclear targets; large aperture magnetic
spectrometer; multicell threshold Cherenkov
counter. For the analyses we use data obtained in the
70 GeV proton beam of IHEP accelerator at
intensity $\approx (5 \div 6) \cdot 10^5$ protons/cycle. The total
statistics of $5\cdot10^7$ inelastic events was collected.  The
sensivity of this experiment for inelastic $pN$-interactions taking in
account the triggering efficiency was $1.6~nb^{-1}$.

\section{Evidence for a baryonic state decaying to $K^0_s p$}

SVD-2 has performed the searches of $\Theta^+$-baryon in two
independent samples of the data selected by the $K^0_s$ decay point.

\begin{figure}[h] \centerline{\psfig{file=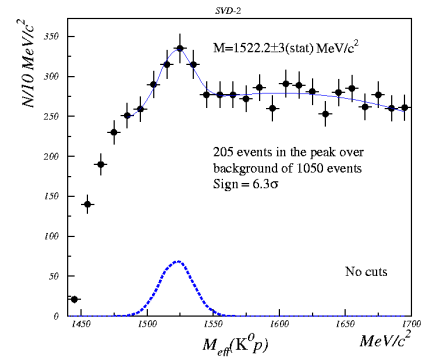,width=2.4in}}
\caption{Analysis I: The $(pK^0_s)$ invariant mass spectrum for
$K^0_s$ decaying inside the vertex detector with the cuts explained in
text.} \label{fig1} \end{figure}

In the first one $K^0_s$ decayed within the vertex detector
were used.  The first step in the analysis was to find out events with
a well defined secondary vertex at the distance of 0 to 35 mm by the
beam direction from the beginning of the active target (corresponding
to the sensitive area of the vertex detector). Tracks from secondary
vertex were explored to the magnetic spectrometer and their momenta
were reconstructed. The $K^0_s p$ invariant mass spectrum
(Fig.\ref{fig1}) shows a clear peak at the $M=1522\pm 3(stat.) \
MeV/c^2$ with the significance of about 6.2$\sigma$ (205 signal over
1050 background events). To estimate the natural width of the observed
peak additional quality cuts were used: 1) the distance of the closest
approach between $V_0$ tracks $\le$ 3 standard deviations and 2)
the number of hits on the track in the magnetic spectrometer $\ge$ 12
(of 18 present). Taking into account the experimental resolution of
the SVD-2 detector (calculated using well-known resonances) it was
estimated that intrinsic width of the $(pK^0_s)$-resonance is
$\Gamma < 14~MeV/c^2$ at 95\% C.L.

\begin{figure}[h] \centerline{\psfig{file=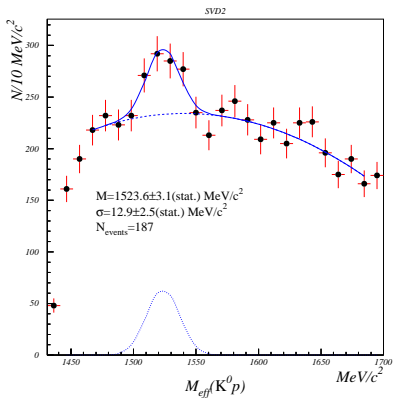,width=2.2in}}
\caption{Analysis II: The $(pK^0_s)$ invariant mass spectrum along
with different background descriptions represented by fitted dashed
histogram: a) mixed-event model background; b) RQMD Monte Carlo
background.} \label{fig2} \end{figure}

For analysis II, the "distant" $K^0_s$ (decay region is $35 - 600\
mm$, outside the vertex detector) were used.
$K^0_s$-mesons were identified by their charged decay mode
$K^0_s \rightarrow \pi^+ \pi^- $.  To eliminate contamination from
$\Lambda$ decays, candidates with ($p \pi^-$) mass hypothesis giving
less than $1.12\ GeV$ were rejected.  About $52000$ $K^0_s$-mesons from
the mass window $\pm 20\ MeV/c^2$ were found in the selected events.
Proton candidates were selected as
positively charged tracks with a number of spectrometer hits more than
12 with a momentum $8\ GeV/c \le P_p \le 15\ GeV/c$. Pions of such
energies should leave a hit in the Cherenkov counter, therefore
absence of hits in TCC was also required. Effective mass of the $K^0_s
p$ system is plotted in Fig.\ref{fig2}. An enhancement is seen at the
mass $M=1523.6\pm 3.1\ MeV/c^2$ with a $\sigma=12.9\pm 2.5\ MeV/c^2$
with statistical significance of 5.6$\sigma$. It was verified that
observed $K^0_s p$-resonance can not be a reflection from other (for
example $K^{*\pm}(892)~or~\Delta^0$) resonances. The mechanism for
producing spurious peak around $1.54~MeV/c^2$ involving $K^0_s$ and
$\Lambda$ decays overlap was also checked. The events were scanned
using SVD-2 event display and no "ghost" tracks were found. It is
impossible to determine the strangeness of this state in such
inclusive reaction, however we did not observe any narrow structure in
$(\Lambda \pi^+)$ invariant mass spectrum in $1500\div1550~MeV/c^2$
mass area (Fig.\ref{fig4}a). When applying a
different cut on $\Lambda$ momentum, $p_{\Lambda}<6~GeV$, we observe a structure near $1480~MeV/c^2$ (Fig.\ref{fig4}b). This peak may correspond to the $\Sigma(1480)$, marked as one-star resonance in the PDG review\cite{pdg}.

\begin{figure}[h] \centerline{\psfig{file=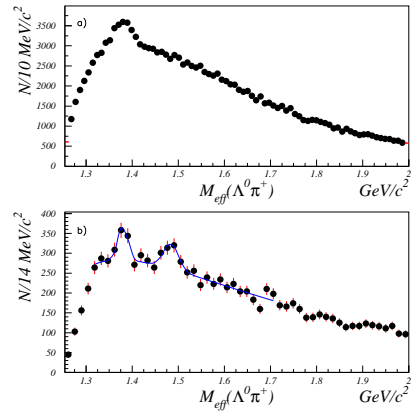,width=2.4in}}
\caption{a) The $(\Lambda \pi^+)$ invariant mass spectrum with the
cuts used for Analysis II. b) The $(\Lambda \pi^+)$ invariant mass
spectrum with the additional cut explained in text} \label{fig5}
\end{figure}

\section{The further analysis of the resonance properties}

\begin{figure}[h] \centerline{\psfig{file=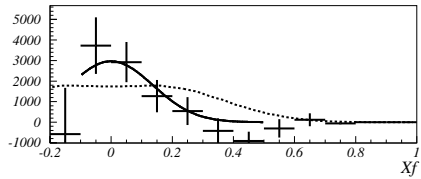,width=2.4in}}
\caption{$X_F$ distribution for the Analysis I.
  Crosses are experimental data,
  normalized curves stand for a quark fragmentation(solid line)
  and a sum of quark and diquark fragmentation(dotted line)
  in the Baranov Regge-based model.
  Y-axis: acceptance-corrected number of events over 0.1 in $X_F$}
\label{fig3} \end{figure}

\begin{figure}[h] \centerline{\psfig{file=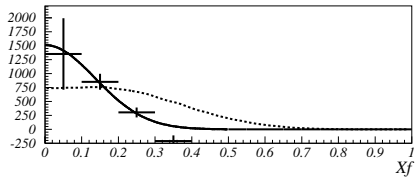,width=2.4in}}
\caption{The same as in fig.\ref{fig3} for the Analysis II.}
\label{fig4} \end{figure}

Our search for $\Theta^+$-particle is an inclusive experiment with a
significant background contribution. We have made an attempt to apply
a subtraction method to investigate $\theta$ creation region in terms
of $X_F$.  An effective mass distribution in a peak region was fitted
with a sum of background (B) and Gaussian functions. Background was
taken as a product of a threshold function and a second degree
polinomial, \mbox{$B(m)=P2(m)(1-e^{-p_2(m-p_1)})$}. All the fit
parameters were given reasonable seeds but no boundaries, to prove a
fit stability. We have chosen a peak region as a gaussian mean $\pm$ 2
gaussian $\sigma$. A number of effective background events under the
peak, $NB_{peak}$, was evaluated by integrating background function
over the peak region. $X_F$ distributions were plotted separately for
a peak and an out-of-peak regions(``wings''); the latter ranged from
the threshold to the 1.7 GeV with a peak region cut out, and a result
was scaled to a $NB_{peak}$. Assuming that the background
characteristics are uniform, we subtract ``wings'' distribution from
the peak one. Choosing more narrow ``wings'' does not change the
general shape of distribution (a rise at $X_F=0$) but shows larger
fluctuations. These operations were performed over the data from both
analyses. Acceptance corrections were taken from the simulations and
were specific for each type of analysis. The results are shown at
figs.\ref{fig3} and \ref{fig4}. We plot also normalized curves of the
predictions made in a Baranov Regge-based model\cite{baranov}. In this
model, overall distribution comes from the sum of quark
fragmentation(bell-shaped at $X_F=0$) and diquark one(seagull-shaped),
taken with somewhat arbitrary weights. In \cite{baranov} the weights
are taken as 1:10, as coming from the analysis of non-exotic baryons
creation. Our data may indicate some favoring to the quark
fragmentation part of the model.

\begin{figure}[h] \centerline{\psfig{file=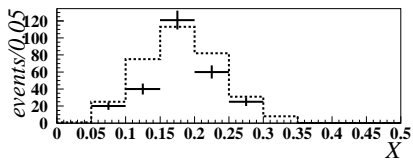,width=2.4in}}
\caption{A projection of an Analysis I result (dotted histogram) onto
Analysis II (crosses).} \label{fig6} \end{figure}

Note that the Analysis II has a specific
narrow acceptance in X due to the proton momentum restrictions. We
"projected" the $X_F$-result of Analysis I to a second one to check
the consistence of our observations. For that, the {\it inverse}
acceptance correction for Analysis II was applied to the result of
Analysis I. We found a plausible agreement of the distribution shapes
(fig.\ref{fig6}) and some difference in a total number of events. The
latter makes a contribution to the cross section error calculations.

As positive as negative results on the theta particle search are
very often presented as the ratio of theta to $\Lambda (1520)$ cross
sections. In our case, we estimate it as 0.04. However we should note
that these two particles may have quite different creation
mechanisms. For example, extracting $X_F$ for $\Lambda (1520)$ the same
way as described above resulted in much more forward-oriented distribution
(not shown here), than for theta particle. It weakens the
reasons of making this comparison. One may suppose that a best
way to present such a ratio would be to accompany it with
corresponding acceptances.

The re-evaluation of the theta creation cross section gives
$\sigma*BR(\Theta \rightarrow K^0p)=6 \pm 3$ $\mu b/nucleon$ for
$X_F>0$. It differs significantly from our previous
estimates\cite{svd4}. The main reason is an unusual form of $X_F$
distribution, assumed flat in our first publications.  Taking into
account our situation of using inclusive data, and undiscovered yet
mechanism of theta particle creation (that means difficulties in
acceptance evaluation), we believe that the cross section value is
subject to future investigations.

In conclusion, SVD-2 observes in two independent samples a signal in
$pK^0_s$ mass distribution with $M=(1523\pm 2 \pm 3)
MeV/c^2\/$,$\Gamma < 14 MeV/c^2\/$, significance of $\approx 8\sigma$,
$\sigma*BR(\Theta \rightarrow K^0p)=6 \pm 3$ $\mu b/nucleon$ for
$X_F>0$. $X_F$ peaks at zero with \mbox{$<|X_F|> \approx 0.1$}, that
agrees qualitatively to a Regge-based model suggested by
Baranov\cite{baranov}. While in agreement with evidences of $\Theta^+$
observation, there is no direct contradiction to null results in
hadron-hadron fixed target collisions, mainly due to different
acceptances at $X_F \approx 0$.

\section{Acknowledgements}

We thank ICHEP'06 Organizing Committee for creating an excellent
scientific atmosphere during the Conference. S.P.Baranov kindly
supplied us with the data used in his work\cite{baranov}, the
discussions with him were also very useful.

\end{document}